\documentclass[10pt,journal,cspaper,compsoc]{IEEEtran}

\usepackage{amsfonts}
\usepackage{amsthm}
\usepackage{graphicx,epsfig,amssymb,amsmath,url,latexsym}
\usepackage[lined, boxed, linesnumbered, commentsnumbered, ruled]{algorithm2e}
\usepackage{subfigure}
\usepackage[square,sort&compress,numbers]{natbib}

\newtheorem{theorem}{Theorem}
\newtheorem{lemma}{Lemma}

\newtheorem{definition}{Definition}

\newcommand{\figsize} {3.5in}

\begin{document}

\title{Moving Window Network Coding in Cooperative Multicast}

\author{\IEEEauthorblockN{Fei Wu,~\IEEEmembership{Student Member,~IEEE}, Cunqing Hua,~\IEEEmembership{Member,~IEEE}, Hangguan Shan,~\IEEEmembership{Member,~IEEE}, Aiping Huang,~\IEEEmembership{Senior Member,~IEEE}}\thanks{This work was supported by the Fundamental Research Funds for the Central Universities (No. 2011QNA5018), Zhejiang Provincial Natural Science Foundation of China (No. LY12F01021), National Natural Science Foundation of China (No. 61001096) and National Basic Research Program of China (No. 2010CB731803). An earlier version of this paper appeared in IEEE Globecom'12 \cite{Wu2012Globecom}.} \thanks{Cunqing Hua is with the School of Information
Security Engineering, Shanghai Jiao Tong University, Shanghai
200240, China. (e-mail: cqhua@sjtu.edu.cn).}\thanks{ Fei Wu,
Hangguan Shan and Aiping Huang are with the Institute of Information
and Communication Engineering, Zhejiang University, China and
Zhejiang Provincial Key Laboratory of Information Network
Technology, Hangzhou 310027, China(e-mails:\{wuff,  hshan,
aiping.huang\}@zju.edu.cn).}\thanks{Correspondence author: Cunqing
Hua}}

\IEEEcompsoctitleabstractindextext{
\begin{abstract}
Cooperative multicast is an effective solution to address the bottleneck
problem of single-hop broadcast in wireless networks. By
incorporating with the random linear network coding technique, the
existing schemes can reduce the retransmission overhead
significantly. However, the receivers may incur large decoding delay and complexity due to the batch decoding scheme. In addition, the dependency on the explicit feedback leads
to scalability problem in larger networks. In this paper, a
cooperative multicast protocol named MWNCast is proposed based on a
novel moving window network coding technique. We prove three
properties of the proposed scheme. Firstly, without explicit
feedback, MWNCast can approach the cooperative capacity with the
packet loss probability dropping almost exponentially with the
increase of window size. Secondly, the average decoding delay of a
receiver is on the order of
$O(\frac{1}{(1-\rho)^2})$ with respect to its traffic intensity $\rho$. Thirdly, MWNCast can achieve the linear decoding complexity of $O(W)$ with respect to the window size $W$. Simulation results show that MWNCast
outperforms the existing schemes by achieving better tradeoff
between the throughput and decoding delay, meanwhile keeping the
packet loss probability and decoding complexity at a very low level
without explicit feedback.
\end{abstract}
}
\maketitle

\section{Introduction}\label{sec:introduction}
Due to the broadcast nature of wireless channels, wireless networks
have been deemed as an efficient solution for multicast file
delivery, multimedia streaming services, etc. Under perfect channel
conditions, multiple clients within the transmission range of a
single transmitter node can receive the same piece of data
simultaneously without incurring any extra overhead. However, this
assumption is invalid in practice since  wireless channels are
subject to fast fading due to signal
attenuation, shadowing and multipath effects, leading to random
failure of packet reception at different clients.

Although packet error can be tolerated to some extents in most
multimedia streaming applications, excessive packet losses are
unacceptable because it can lead to the degradation of quality of
experience (QoE) to the end users. In order to improve the
reliability of multicast, many techniques and protocols have been
developed. One class of solutions follow the error recovery path
that tries to tackle the packet loss problem using the automatic
repeat request (ARQ) or combined with forward error correction (FEC)
(e.g., \cite{2006two,zhang2006}), which however lead to feedback
storm problem since the source node relies on the feedback from
clients to make retransmission decisions. To address this issue,
another class of schemes adopt the rateless coding strategy (e.g.,
\cite{luby2002,Shokrollahi2006,ho2004}), whereby the source node
keeps transmitting coded symbols without explicit feedback, and any
clients can decode the packet after accumulating enough symbols.
Although such approaches are able to provide reliable transmissions,
they may suffer from the bottleneck problem, that is, the throughput
of the overall system is limited by the node with the worst channel
capacity.

As a natural solution to the bottleneck problem in multicast,
cooperative communications have drawn increasing attentions
recently. In \cite{alay2009}, integrated with layered video coding
and packet level forward error correction, the randomized
distributed space time codes are adopted to design cooperative
multicast scheme that can provide efficient and robust video
delivery. Relay selection has been studied in \cite{rong2010} to
improve the performance of cooperative multicast in a mobile
computing environment. The outage probability with cooperative
multicast is analyzed in \cite{zhao2010}, which suggests that the
performance can be improved with more relay nodes. These
schemes demonstrate the effectiveness of physical-layer cooperation
in alleviating the bottleneck problem in multicast, but they may
incur some difficulties in practical implementation, such as tight
time synchronization. Furthermore, the sequential retransmissions of
the lost packets to multiple receivers (requested by feedback) can
reduce the bandwidth efficiency.

One potential way to address this issue is to utilize network coding
techniques whereby the lost packets can be encoded together to
reduce the number of retransmissions. For example, \cite{fan2009}
shows the benefit of cooperation at the network layer via a simple
XOR network coding technique. In \cite{jin2009}, the random linear
network coding (RLNC) \cite{ho2004} is adopted for multicast
applications, and the channel and power allocation in relaying nodes
are optimized for maximizing the multicast rate. It is shown in
\cite{fanous2010} that compared to the physical-layer cooperation,
the use of RLNC at the relays can enhance the system throughput. In
\cite{Pac}, a RLNC-based opportunistic multicast protocol is
proposed which can alleviate the bottleneck problem effectively.
However, to avoid throughput degradation, the block size in RLNC has
to scale with the the number of receivers \cite{Swapna2010}, which
in turn leads to large decoding delay and complexity. In addition,
the centralized scheduling policies in
\cite{fan2009,jin2009,fanous2010} rely on the feedback from the
relays and receivers about the packet reception status, which make
them difficult to scale to larger network size in practice.

In this paper, a cooperative multicast protocol named MWNCast is
proposed based on the moving window network coding (MWNC) technique.
By exploiting the residual capacity of relay nodes to serve the
bandwidth starving receivers, the proposed scheme can effectively
alleviate the bottleneck problem in wireless multicast. Based on the
random walk and point process theory, we prove three fundamental
properties of MWNCast. Firstly, without explicit feedback, MWNCast
can approach the cooperative capacity with the packet loss
probability dropping almost exponentially with the increase of
window size. Secondly, if the coding window is large enough such
that the packet loss can be neglected , the average decoding delay
experienced by a receiver is $O(\frac{1}{(1-\rho)^2})$, where $\rho$
is the traffic intensity of the node. Moreover, the decoding delay of
different receivers are mutually independent, which can guarantee
the scalability of the scheme in large networks. Thirdly, MWNCast
can achieve the minimal decoding complexity $O(W)$ (W is coding
window size) for a given target throughput. We provide simulation
results to validate the theoretical results, which show that the
proposed scheme not only can guarantee reliable transmission without
explicit feedback, but also can achieve high throughput with reduced
decoding delay and complexity.

The rest of this paper is organized as follows. In
Section~\ref{sec:system_model}, the system models assumed in this
paper is introduced. We present MWNC in Section~\ref{sec:MWNC}. In
Section~\ref{sec:design}, an overview of MWNCast is firstly
provided, followed by its functional modules in detail. In
Section~\ref{sec:analysis}, we establish the theoretical framework
and then prove three key properties of MWNCast. Simulation results
are provided in Section~\ref{sec:simulation} and finally we conclude
this paper in Section~\ref{sec:conclusion}.

\section{System Model}\label{sec:system_model}
We consider a wireless network consisting of a source node $s$ and a
set of $\mathcal{N}$ receivers. The source node has a stream of
packets to be transmitted to all receivers. As discussed in previous
section, for lossy wireless networks, the capacity of plain
broadcast (even with a sophisticated network coding scheme) is
limited by the worst receiver. To address this problem, we adopt a
cooperative networking structure, whereby a subset $\mathcal{R}
\subseteq \mathcal{N}$  of nodes are selected as \emph{relays },
which perform not only the normal receiving function to receive data
from the source, but also the {\em relaying} function that forwards
the received data to the remaining subset $\mathcal{E}$ of {\it end
receivers} ($\mathcal{E} = \mathcal{N}\backslash \mathcal{R}$). To
simplify the design of protocol, we assume that relay nodes only
receive data from the source, while the {\it end receivers} can
receive data from both the source and the relay nodes.

Similar to \cite{jin2009}, we assume that there are $K$ orthogonal
channels that can be operated by each node\footnote{We assume
frequency division multiple access (FDMA) in this paper, but it can
be easily generalized to time division multiple access (TDMA) too.}.
Therefore, in order to avoid co-channel transmission interference
between the source and the relay nodes, at most $K-1$ relay nodes
are allowed to transmit concurrently with the source. Time is
divided into slots, and each node is equipped with one half-duplex
radio, so a relay node cannot receive and relay at the same time.

To characterize the lossy nature of the wireless channel, let
$C_{i,j}$ denote the packet reception probability (PRP) for a pair
of nodes $i$ and $j$ \cite{More}. In this paper, we assume the PRPs of all links in the network are quasi-static and
collected by the source node through some online or offline
measurements \cite{Padhye2005}\cite{Reis2006}. Note that $C_{i,j}$
is equivalent to the capacity of link $(i,j)$ since it is the
maximum achievable throughput for error-free transmission from node
$i$ to $j$. In the following, without abusing the notation, we refer
to $C_{i,j}$ the PRP as well as the capacity of the link. In
particular, let $C_{0,j}$ denote the link capacity from the source
to any node $j\in \mathcal{N}$.

\section{Moving Window Network Coding}\label{sec:MWNC}
To improve the performance of wireless multicast, many different
network coding techniques have been proposed  from different
perspectives. The random linear network coding scheme \cite{ho2004}
adopts a block transmission strategy which can approach the capacity
with less feedback overhead. Unfortunately, it is shown in
\cite{Swapna2010} that the block size of RLNC has to scale with the
increase of the number of receivers to avoid the loss of throughput,
which however will result in large decoding delay. The ARQ-based
online network coding (ANC) \cite{Kumar2008} achieves the one-hop
maximum multicast throughput, but the decoding delay of the
receivers with worse channel conditions is unfairly large. Many
solutions have been proposed for this problem (e.g.,
\cite{2009minimizing,2008online,Barros2009,sorour2010}. When the
number of receivers is small, the schemes proposed in
\cite{2009minimizing} and \cite{2008online} can reduce the decoding
delay, but the optimal throughput and decoding delay cannot be
achieved simultaneously for larger network size \cite{keller2008}.
In \cite{Barros2009}, a delay threshold based on scheme is proposed
to incorporate with the ANC scheme, which can guarantee the decoding
delay to be within the prescribed bound at the cost of throughput
degradation. The instant decodable network coding can effectively
minimize the decoding delay, but it cannot guarantee the order of
decoding \cite{sorour2010}. Note that most of these delay control
schemes rely on the feedback from receivers. With virtually no
feedback information, the optimal RLNC strategy for
delay-constrained traffic is studied in \cite{hou2011}, but the
scheme still suffers from the throughput degradation problem of RLNC
with the network scale increase.

\begin{figure}[t]
\centering
  \includegraphics[width=2.5in]{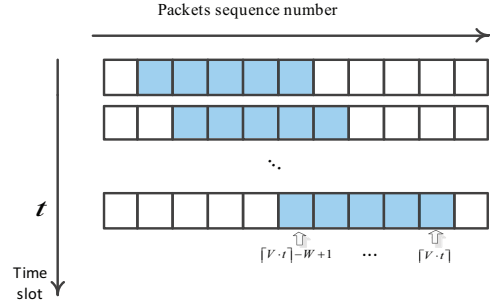}
  \caption{Encoding of MWNC.}
  \label{fig:encoding}
\end{figure}

\begin{figure}[t]
\centering \subfigure[Received symbols]{
 \includegraphics[height=1.5in]{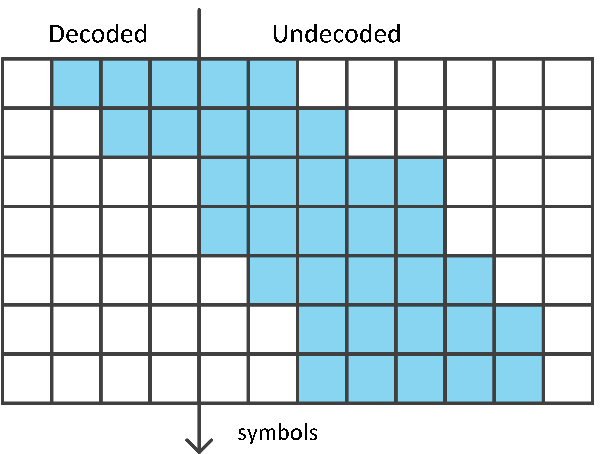}}
 \hspace{0.2in}
\subfigure[After Gauss-Jordan elimination]{
  \includegraphics[height=1.5in]{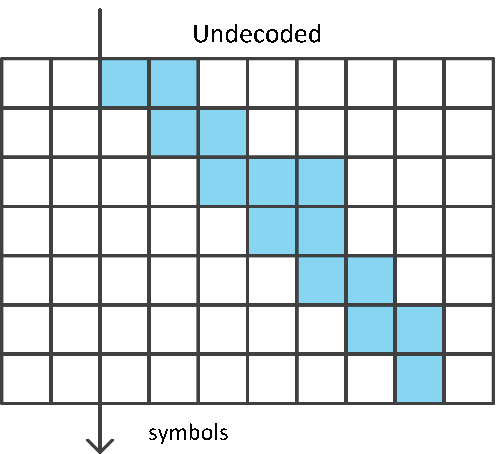}}
  \caption{Decoding of MWNC.}
  \label{fig:how decode}
\end{figure}

Motivated by these techniques,  we propose the MWNC scheme to combine the advantageous features of
traditional network coding schemes\cite{Wu2012Pimrc}. 
MWNC adopts the encoding
strategy similar to RLNC, but the block of packets to be encoded in
each slot is moving forward at a constant speed $V$ (see
Fig.~\ref{fig:encoding}). Specifically, at time slot $t$, a block of
$W$ packets with the sequence number ranging from $\lceil V\cdot t
\rceil -W +1$ to $\lceil V\cdot t \rceil$ are encoded with random
coefficients on a finite field, which are also transmitted with the
coded symbol.\footnote{$\lceil \cdot \rceil$ is the ceil function to
guarantee that the boundaries of the window are aligned to integer
values. Note that if $\lceil V\cdot t \rceil < W$, then the block is
started from $1$ to $\lceil V\cdot t \rceil$.} After overhearing the
coded symbols from the source, the receiver attempts to decode the
original packets through Gauss-Jordan elimination approach. A
typical example of the decoding process is shown in Fig.
\ref{fig:how decode}, in which the Gauss-Jordan Elimination can be
performed progressively as the coded symbol arrives and finally the
original packets can be retrieved when the reduced matrix has full
rank (Fig. \ref{fig:how decode}(b)). Note that $V$ represents the
target throughput, so it should be within the network capacity.


In Table~\ref{tab:MWNC}, we show an example where the window size
$W=3$ and the moving speed $V=0.5$. In this example, the source
starts by sending the uncoded packets $p_1$ twice in the first two time
slots, one of which is lost by the receiver. Then it sends coded
symbols $p_1\oplus p_2$ with randomly chosen coefficients in the
next two slots (since $\lceil Vt \rceil = 2$ for $t=3,4$), one of
which gets received, so the receiver can successfully decode $p_2$
at the forth time slot. From the fifth time slot, a full window of
three packets are encoded in each time slot, which is moved forward
with the speed of $V=0.5$. Note that because there is no feedback
mechanism in MWNC, it cannot guarantee $100\%$ reliability. For
example, $p_3$ will get lost at the $12^{th}$ time slot, since it
will never be decoded after the window has moved to $p_5,p_6,p_7$,
even when the client has received the information of $p_3$ and $p_4$
at the $8^{th}$ time slot. However, we will prove in Section
\ref{sec:analysis} that the packet loss probability with MWNC drops
almost exponentially with the increase of window size.

MWNC has some other interesting properties. Firstly, the decoding
opportunity exists in each time slot, therefore it avoids the
intrinsic decoding delay problem incurred by RLNC. In addition, the
decoding opportunity is balanced between clients with good and poor
channel conditions, so none of the clients will be dominated by
other clients with better channel conditions. Secondly, the coding
coefficient matrix in buffer is very sparse (see Fig. \ref{fig:how
decode}) due to the moving window strategy, so the decoding
complexity of MWNC is much lower than RLNC. In
Section \ref{sec:analysis}, we will develop some theoretical models
to analyze these properties.

\begin{table}[t]
\caption{Example of Moving Window Network Coding ($W=3, V=0.5$)}
\vspace{-0.2cm} \label{tab:MWNC} \centering
\begin{tabular}{c|c|c|c|c}
  \hline
  Time & $\lceil Vt \rceil$ & Sent symbols & Received? & Decoded/Lost \\
  \hline
  1 & 1 & $p_1$             & $\times$  & --  \\
  2 & 1 & $p_1$   & $\checkmark$ & D: $p_1$ \\
  \hline
  3 & 2 & $p_1\oplus p_2$   & $\times$ & --  \\
  4 & 2 &  $p_1\oplus p_2$ & $\checkmark$ & D: $p_2$ \\
  \hline
  5 & 3 & $p_1\oplus p_2\oplus p_3$ & $\times$ &-- \\
  6 & 3 & $p_1\oplus p_2\oplus p_3$ & $\times$ & --  \\
  7 & 4 & $p_2\oplus p_3\oplus p_4$ & $\times$ &-- \\
  8 & 4 & $p_2\oplus p_3\oplus p_4$ & $\checkmark$ &--  \\
  9 & 5 & $p_3\oplus p_4\oplus p_5$ & $\times$ &-- \\
  10 & 5 & $p_3\oplus p_4\oplus p_5$ & $\times$ & --\\
  11 & 6 & $p_4\oplus p_5 \oplus p_6$ & $\times$ &-- \\
  12 & 6 & $p_4\oplus p_5 \oplus p_6$ & $\times$ & L: $p_3$,$p_4$ \\
  \hline
  13 & 7 & $p_5\oplus p_6 \oplus p_7$ & $\checkmark$ &-- \\
\end{tabular}
\vspace{-0.0cm}
\end{table}

Note that the concept of network coding over a moving window has been
considered in \cite{sundararajan2009network} and
\cite{lin2010slideor}. In \cite{sundararajan2009network}, RLNC is
incorporated with the congestion window in TCP protocol to improve 
the throughput in the lossy wireless environment. In
\cite{lin2010slideor}, SlideOR is proposed to encode packets in
overlapping window, which can avoid the throughput loss in
opportunistic routing. Our scheme differs from these schemes in the following aspects. Firstly, MWNC can achieve better control of the decoding delay and complexity with appropriate settings of the moving speed and window size, while these schemes are best-effort and there is no guarantee for the decoding delay at the receivers. In addition, we develop theoretical models to characterize the reliability, decoding delay and
decoding complexity properties of MWNC. Secondly,
these schemes rely on the feedback of the receivers to move forward the coding window, which is nontrivial in wireless broadcasting since the ACKs of different receivers have to be carefully scheduled to avoid collision.
In addition, even if the reliability of ACKs can be guaranteed, the feedback delay may lead to the degradation of the network throughput\cite{feedback}. In our scheme, the coding window is moved forward according to a prescribed moving speed $V$, which does not rely on the feedback from the receivers. Of course, $V$ should be carefully set to be within the network capacity to avoid overwhelming the receivers, which is not difficult since the link capacity is assumed to be quasi-static. If the network is dynamic, this parameter should be adapted according to the network condition, which however is beyond the scope of this paper.

\section{Design of MWNCast}\label{sec:design}
In this section, we propose MWNCast, a cooperative multicast
protocol based on the MWNC technique. Before elaborating on the
details of the protocol, we briefly introduce the motivation and
basic functionality of MWNCast with a simple example.

\subsection{Overview of MWNCast}
\begin{figure}[t]
\centering \subfigure[Topology]{
 \includegraphics[height=2in]{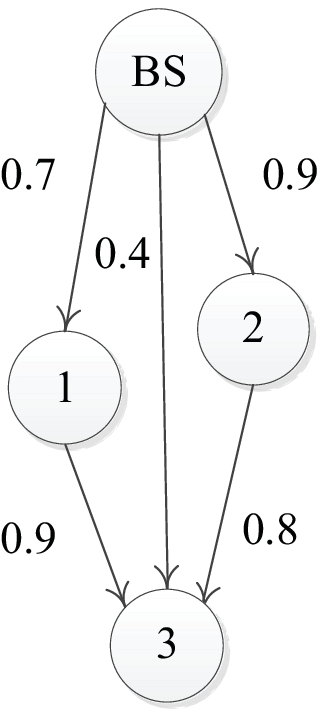}}
 \hspace{0.2in}
\subfigure[MWNCast]{
  \includegraphics[height=2in]{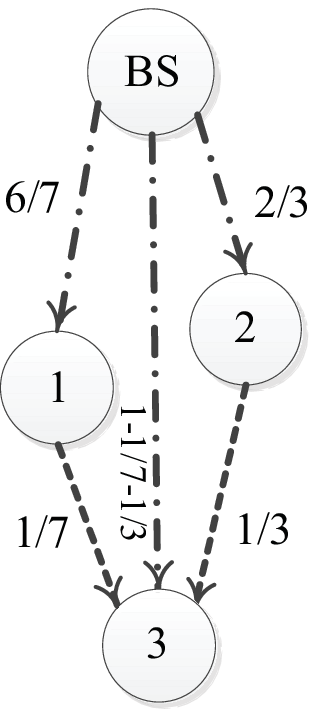}}
   \hspace{0.2in}
\subfigure[TX scheduling]{
  \includegraphics[height=2in]{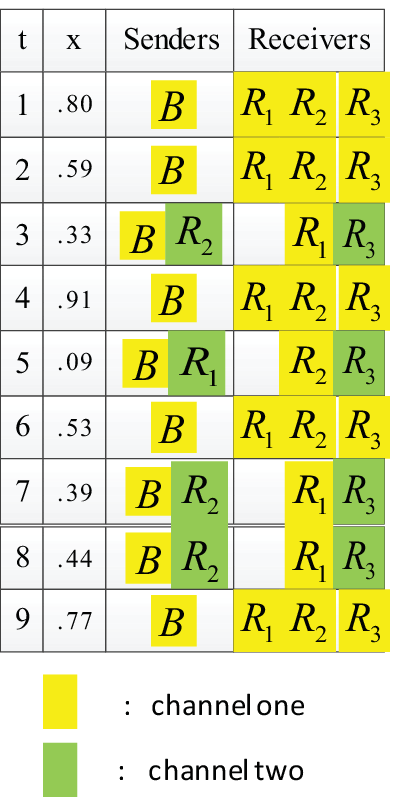}}
   \hspace{0.2in}
\subfigure[Encoding]{
  \includegraphics[height=2in]{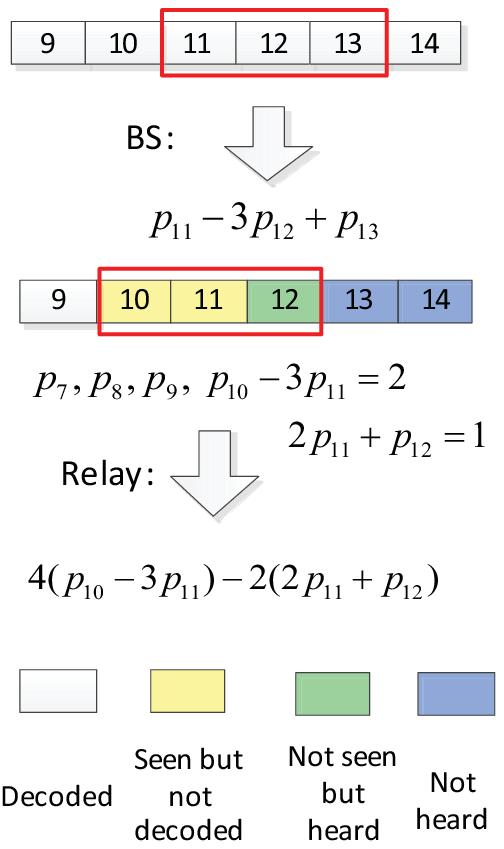}}
  \caption{A simple example of MWNCast.}
  \label{fig:example}
\end{figure}

Consider a simple example as shown in Fig. \ref{fig:example}(a),
which consists of three receivers, the number on each link is the
corresponding PRP. For plain broadcast, it is easy to see the
capacity of the system is $0.4$ due to the bottleneck receiver
$R_3$, which requires more time to receive the same amount of
information as that of clients $R_1$ and $R_2$. Therefore, some time
is wasted for clients $R_1$ and $R_2$ since the information sent by
the source is not innovative to these two receivers after they have
received the required data. On the other hand, if these two clients
have packets that are not received by client $R_3$, one of them can
forward the packets to client $R_3$ on behalf of the source on a
different channel using its residual time, while the other client
can continue receiving data from the source. Ideally, if clients
$R_1,R_2$ are assigned to devote $1/7$ and $1/3$ of their time to
serve client $R_3$ alternately while spending the rest of their time
to receive from the BS, then the maximum achievable throughput for
$R_1$ is $0.7 \times \frac{6}{7}=0.6$, and $R_2$ is $0.9 \times
\frac{2}{3}=0.6$. Meanwhile, client $R_3$ can receive data
alternately from clients $R1,R_2$ when they are active, and from the
BS in the rest time, so its achievable throughput is
$\frac{1}{7}\times 0.9+\frac{1}{3}\times
0.8+(1-\frac{1}{7}-\frac{1}{3})\times 0.4>0.6$ (see Fig.
\ref{fig:example}(b)), which suggests that the throughput of $0.6$
(packet/slot) can be achieved through this cooperation scheme.

The key to the success of this cooperative strategy is the scheduling of
the relay transmissions, that is, to determine which set of relay
node should transmit at a specific time slot. To this end, we adopt
a stochastic scheduling method, which works as follows. At the
beginning of each time slot, the source generates a random variable
$x$ uniformly distributed in $[0,1]$. If $0\le x<\frac{1}{7}$, then
$R_1$ is selected to relay the data to $R_3$, while $R_2$ keeps
receiving from the source. If $\frac{1}{7}\le
x<\frac{1}{7}+\frac{1}{3}$, the roles of $R_1$ and $R_2$ are
exchanged. Otherwise, only the source transmits and all clients
receive information from it. This scheduling decision is broadcasted
to all relays. Each second-hop receiver always receives from the
best transmitter (the source or a relay). An example of the
scheduling sequences is shown in Fig. \ref{fig:example}(c).

The source and the selected relays will transmit at the scheduled time slot. The packets to be transmitted are encoded using the MWNC technique, which range from $\lceil V\cdot t \rceil -W +1$ to $\lceil
V\cdot t \rceil$ at time $t$. The source transmits the encoded symbol on its channel, while the selected relay node transmits a specific encoded symbol on a different channel. For relay node, this encoded symbol is generated from a batch of
packets (the most close to the expected window in the relay's
buffer), including the newly decoded packets and the combination of
undecoded packets (e.g., $p_{10}$, $p_{11}$, and $p_{12}$ in Fig.
\ref{fig:example}(d)).  After overhearing the transmissions from the
source and the relay, the end receiver attempt to decode
the original packets with Gauss-Jordan elimination technique.



\subsection{MWNCast Protocol}
In this subsection, we discuss the details of MWNCast protocol. We
have explained how to implement MWNC in a cooperative scenario, so
in the following we focus on the cooperative scheduling in MWNCast,
which can be decomposed into three modules, namely the selection of
relay nodes, the allocation of relay time, and the online scheduling
of relay transmissions.

\subsubsection{Relay node selection}
The ultimate goal of MWNCast is to alleviate the bottleneck and
maximize the multicast capacity through the cooperation of relays.
To this end, we propose a procedure to search for a set of candidate
relay nodes that can achieve the maximum capacity. The basic idea is
as follows. For a given target network capacity $C_T$, we can
partition the set of $\mathcal{N}$ receivers into two groups
according to the PRPs from the source to these nodes, then the nodes
with PRPs above $C_T$ will be selected as the relay nodes since
their residual capacities can be used for serving the remaining end
receivers.

The rational for selecting relay nodes in this way is that the
selected relay nodes can meet the target capacity requirement.
However, it cannot guarantee that the remaining nodes can achieve
the target capacity $C_T$ as well, since their achievable throughput
depends on their link capacities to the relay nodes, as well as how
much residual time of these relay nodes can be devoted for
cooperation. Therefore, it is necessary to check the feasibility of
this target $C_T$, which involves the computations of the available
cooperation time for a given set of relays (to be discussed), with
which we can compute the achievable throughput for each receivers.
If any of the nodes fail to achieve the target capacity $C_T$, then
it means that this target capacity is infeasible and a smaller value
should be attempted, otherwise a larger capacity can be supported.

Based on this idea, we propose a binary search procedure to find the
maximum achievable capacity as shown in
Algorithm~\ref{alg:relay_node_search}. The algorithm maintains a
lower threshold $C_L$ and an upper threshold $C_U$ for the target
capacity initially. Then starting with $C_T=(C_{L}+C_{U})/2$, a set
of relay nodes with qualified link capacities are determined (lines
5-6). The achievable capacities of the remaining nodes are computed
using Algorithm 2 (line 7). If the target capacity can be achieved
by all nodes, then the lower threshold is increased to $C_T$ (line
9), otherwise the upper threshold is reduced to $C_T$ (line 11). The
same procedure is repeated until the upper and lower thresholds
converge. Finally, the algorithm returns the set of qualified relay
nodes for the maximum achievable capacity $C_T^*$.

\begin{algorithm}[t]
\SetLine \Begin{
  $C_L \leftarrow 0, C_U\leftarrow 1$;

  \While {$C_U-C_L>\Delta$} {
  $C_T \leftarrow (C_U-C_L)/2$\;

  $\mathcal{R} \leftarrow \{j|C_{0,j} \geq C_T, j\in \mathcal{N}\}$\ ;

  $\mathcal{E} \leftarrow \mathcal{N}\backslash \mathcal{R}$\;

  Call Algorithm 2 to check the feasibility of $C_T$ for the {\it relay} set $\mathcal{R}$ and the {\it receiver} set $\mathcal{E}$\;

  \uIf {$C_T$ is feasible} {$C_L \leftarrow C_T$;}
  \Else {$C_U \leftarrow C_T$;}
  }
} \caption{Relay node selection} \label{alg:relay_node_search}
\end{algorithm}

\subsubsection{Relay time allocation}
As discussed in the previous subsection, for a target capacity
$C_T$, if a node $i$ has a PRP of $C_{0,i} > C_T$,  it is selected
as a candidate relay node. In this case, at least a $C_T/C_{0,i}$
fraction of its time has to be used for receiving data from the
source so that the target capacity requirement can be satisfied. As
a result, its residual time is at most $(C_{0,i}-C_T)/C_{0,i}$,
which can be used for serving the remaining receiver nodes.
Therefore, the next problem is to find the allocation of the relay
time for each relay node under its residual time budget, such that
the target capacity requirement of the end receiver nodes can also
be satisfied. If such time allocation exists, it means the target
capacity is achievable, and vice versa.

To this end, we propose a relay time allocation algorithm as shown
in Algorithm~\ref{alg:relay_time_allocation}, which proceeds in
round as follows. In the beginning, each candidate relay node $i\in
\mathcal{R}$ is initialized with the residual time $C_i =
(C_{0,i}-C_T)/C_{0,i}$ (line 2), and each end receiver node $j\in
\mathcal{E}$ has a residual capacity demand $D_j = C_T$ (line 3). In
each round $l$, a greedy algorithm (algorithm 3) is invoked to
select a subset $\mathcal{R}_l$ from $\mathcal{R}$ with at most
$K-1$ elements, such that the overall capacity of all receivers in
$\mathcal{E}$ is maximized (line 7). The capacity of a node $j$ is
determined as $C_{R(j),j}$, whereby $R(j)$ is the node in
$\mathcal{R}_l\bigcup\{s\}$ that provides the maximum capacity to
node $j$ among all nodes in $\mathcal{R}_l$.

Given the relay subset $\mathcal{R}_l$, the next step is to decide
the time ratio $\tau_l$ that they can devote for relaying.  Note
that since none of the relay nodes should contribute more than its
residual time, and none of the receiver nodes should get service
more than its residual capacity demand, so the rely time $\tau_l$
for this subset $\mathcal{R}_l$ is set to the minimum of the
residual time of these relay nodes and the residual demand of all
receivers (line 8), then for each selected relay node $i$, its
residual time is reduced by the amount of $\phi_l$ (line 10). If its
residual time is used up, it is removed from the candidate relay set
and will not participate in the relay time allocation in the next
round (line 11). Similarly, for each receiver $j$, its residual
demand is reduced by an amount of $\phi_l*C_{R(j), j}$, which is the
effective throughput it will receive from this set of relay nodes
(line 14). If its demand is satisfied, it is removed from receiver
set and will be considered in the next round (line 15). The same
procedure is repeated to find the next subset of relay nodes and its
relay time allocation, until either the candidate relay set
$\mathcal{R}$ or the receiver set $\mathcal{E}$ becomes empty, or
the overall relay time reaches 1 (line 8). If the receiver set
$\mathcal{R}$ is empty eventually, it means that the target capacity
demand $C_T$ can be met by all receiver nodes, then the algorithm
returns a list of relay subsets and their corresponding relay time;
Otherwise, it means the target capacity $C_T$ is infeasible and the
algorithm returns an empty set.

\begin{algorithm}[t]
\SetLine


\Begin{
  $C_i \leftarrow (C_{0,i}-C_T)/C_{0,i}, \forall i \in \mathcal{R}$\;
  $D_j \leftarrow C_T, \forall j \in \mathcal{E}$\;
  $l \leftarrow 0$ \;
 \While{$\mathcal{R} \neq \emptyset$ and $\mathcal{E} \neq \emptyset$ and $\sum_l \phi_l \leq 1$} {
   $l \leftarrow l+1$\;
   Call Algorithm 3 to select $\mathcal{R}_l \subseteq \mathcal{R}\bigcup\{s\}$ such that $|\mathcal{R}_l|\leq K$ and $\sum_{j\in \mathcal{E}} C_{R(j),j}$ is maximized, where $R(j) \leftarrow \arg \max_{i\in \mathcal{R}_l \bigcup \{s\}} C_{i,j}$\;

   $\phi_l \leftarrow \min\{\min_{i\in \mathcal{R}_l} C_i, \min_{j\in \mathcal{E}} D_j/C_{R(j),j}, 1-\sum \phi_i\}$\;

   \ForEach{$i \in \mathcal{R}_l$} {
        $C_i \leftarrow C_i - \phi_l$;

        \lIf {$C_i \leq 0$} {$\mathcal{R} \leftarrow \mathcal{R} \backslash i$;}
    }
   \ForEach{$j \in \mathcal{E}$}{

    $D_j \leftarrow D_j - \phi_l*C_{R(j),j}$;

    \lIf {$D_j\leq 0$} {$\mathcal{E} \leftarrow \mathcal{E}\backslash j$;}
    }
 }
   \lIf {$\mathcal{E} == \emptyset$} {\KwRet{$\{\mathcal{R}_l, \phi_l\}_{l\in \mathcal{L}}$}}\;
   \lElse {\KwRet{ $\{\emptyset\}$};}
 }

\caption{Relay Time Allocation} \label{alg:relay_time_allocation}
\end{algorithm}

In each round of Algorithm 2 (line 7), we need to find a subset of
at most $K-1$ relay nodes that can provide maximum capacity to the
unsatisfied receivers together with the source node. Let
$\mathcal{B} = \mathcal{R}\bigcup \{s\}$ denote the set nodes
consisting of the candidate relay set $\mathcal{R}$ and the source
node $s$. The capacity of a selection of relay nodes $\mathcal{R}_l
\in \mathcal{B}$ is defined as $C(\mathcal{R}_l) = \sum_{j\in
\mathcal{E}} C_{R(j), j}$. Our objective is to find a selection
$\mathcal{R}_l$ with the maximum capacity such that the
$|\mathcal{R}_l|\le K$. This problem is known as a special case of
the generalized maximum coverage problem, which is NP-hard
\cite{Cohen2008}. To solve this problem, we introduce the following
definitions:

\begin{definition} (residual capacity/weight)
Consider a selection $\mathcal{R}_l$, a relay $i$ and a receiver
$j$. We define the residual capacity $C_{\mathcal{R}_l}(i,j)$ to be
equal to $C_{i,j}-C_{R(j),j}$.
\end{definition}

\begin{definition}(addition of a relay) For a
selection $\mathcal{R}_l$ and a relay $i\not\in \mathcal{R}_l$, we
define $\mathcal{R}_l\oplus i$ as the addition of $i$ to
$\mathcal{R}_l$. In other words, $\mathcal{R}_l\oplus i$ is a new
selection $\mathcal{R'}_l$, and
\begin{eqnarray}
R'(j)=\left\{
\begin{array}{ll}
i, & \mbox{\normalfont if} \: C_{i,j}>C_{R(j),j} \\
R(j), &\mbox{\normalfont otherwise}.
\end{array}
\right.
\end{eqnarray}
\end{definition}

Base on these concepts, we develop a greedy algorithm as shown in
Algorithm~\ref{alg:greedy_coverage}. The basic idea is to
incrementally add the relay node with the maximum positive residual
capacity, so that the overall capacity is non-decreasing. At line 3,
all receivers are initially assigned to the source. Then in each
round, one of the candidate relays that has the maximum positive
residual capacity is selected to join the relay node set
$\mathcal{R}_l$ (line 6) until $|\mathcal{R}_l|$ exceeds $K$. It can
be proved that this greedy algorithm can achieve an approximation
ratio of $1-(1-\frac{1}{K-1})^{K-1}$ to the optimal solution
\cite{Cohen2008}.

\begin{algorithm}[t]
\SetLine \Begin{
  $\mathcal{R}_l \leftarrow \{s\}$\;
  $R(j) \leftarrow s, \forall j\in \mathcal{E}$\;
  \While{$|\mathcal{R}_l|\leq K$}{
    Find a relay $i\in \mathcal{R}$ with the maximum residual capacity, i.e., $i\leftarrow \arg \max_{i'\notin \mathcal{R}_l} \sum_{j\in \mathcal{E}} C_{\mathcal{R}_l}(i', j)$\;
   \lIf{$C_{\mathcal{R}_l}(i, \mathcal{E}) > 0$}{

   $\mathcal{R}_l \leftarrow \mathcal{R}_l \oplus i$\;
   }
   \lElse {break;}
  }
  \KwRet{$\mathcal{R}_l$.}
} \caption{Greedy Maximum Capacity Relay Selection}
\label{alg:greedy_coverage}
\end{algorithm}

\subsubsection{Online Relay Transmission Scheduling}
From Algorithms 1 and 2, we can find a list $\mathcal{L}$ of relay
node set $\mathcal{R}_l$ and the corresponding relay time allocation
$\phi_l$, such that the multicast capacity of the system is
maximized. Let $C^*$ denote the maximum capacity corresponding to
the results, then for any capacity requirement $C\leq C^*$, we
should have:
\begin{equation}\label{eq:scheduling}
 C\leq \sum_{l\in \mathcal{L}} \phi_l*C_{R(j), j}, \forall j \in \mathcal{N}.
\end{equation}

From \eqref{eq:scheduling}, we can see that the amount of time that
a subset $\mathcal{R}_l$ to be scheduled for relaying should be
proportional to $\phi_l$, such that the required capacity can be
satisfied. Since the time is slotted, as briefly introduced in last
section, we can adopt a stochastic online algorithm to approximate
the scheduling. Specifically, let us define $\psi_l$ as
\begin{equation}
\psi_l = \sum_{k\leq l} \phi_l, \forall l.
\end{equation}

In each time slot $t$, the source generates a random number between
0 and 1, if its value falls between $\psi_k$ and $\psi_{k+1}$, then
the $k^{\text{th}}$ subset of relay nodes are selected for relaying
in this time slot. It is easy to see that this stochastic scheduling
policy converges to the required proportional of time for each relay
set in a long run. This schedule algorithm can be executed by the
source in an online fashion at the beginning of each time slot, and
an unique channel is assigned to each selected relay node. The
scheduling results (relay nodes and their operating channels) are
broadcasted to all receivers, then they can choose the best relay
node and switch to the corresponding channel to receive the data.

\section{Analysis}\label{sec:analysis}
In this section, we develop some theoretical models to characterize the basic properties of MWNCast. Firstly, we introduce the
\emph{equivalent channel capacity} model, which is an unified model
for characterizing the capacity of both relay and receiver nodes.
Based on this model, the decoding delay, reliability and decoding complexity properties of MWNCast are analyzed using the random walk and point process
theories.

\subsection{Equivalent Channel Capacity Model}
As discussed in Section~\ref{sec:system_model}, the capacity of a
point-to-point wireless link $(i,j)$ is given by the PRP $C_{i,j}$.
However, the analysis of the link capacity in MWNCast is complicated
since: (i) a relay node may not stay in the ``receiving'' state all
the time; (ii) a receiver node may receive data from different relay
nodes at different time slots. In this subsection, we propose an
\emph{equivalent channel capacity} model to characterize the
capacity of these two kinds of nodes.

For a relay node $i\in \mathcal{R}$, its aggregated fraction of time
in the  ``relaying'' state is given by $\Phi_i = \sum_{l\in
\mathcal{L}, i\in \mathcal{R}_l}\phi_l$.  Since the online relay
scheduling algorithm is a stochastic scheme, we can assume that in
each time slot, the probability for the node to receive from the
source is $1-\Phi_i$, and the probability for relaying is $\Phi_i$.
Taking into account the PRP from the source, we can define the
equivalent channel capacity $\hat{C}_i$ for this relay node as
$\hat{C}_i = (1-\Phi_i)C_{0,i}$, which is the maximum achievable
throughput of this node from the source without errors.

For a receiver node $j$, if a relay subset $\mathcal{R}_l$ is
selected for transmission (with a probability of $\phi_l$), it will
choose to receive from the best relay node $R(j)$ (including the
source) with the maximum PRP, i.e., $R(j)=\arg \max_{i\in
\mathcal{R}_l\bigcup \{s\}} C_{i,j}$. Therefore, we can define the
equivalent channel capacity $\hat{C}_j$ for node $j$ as the
aggregated throughput from all relay subsets, i.e., $\hat{C}_j =
\sum_{l\in \mathcal{L}} \phi_l C_{R(j),j}$.

Note that this equivalent channel capacity model is an approximation
of the link capacity for the two kinds of nodes in MWNCast, which
makes it tractable to analyze the reliability and decoding delay
properties in the following subsections.

\subsection{Preliminary Property of MWNC}

In this subsection, we establish some basic properties for MWNC
using the random walk and point process theories, with which we can
analyze the performance of MWNCast.

In the following, we use the ``packet'' to denote the original data,
and the ``symbol'' to denote the linear combination of the packets
within the window. For any receiver with capacity $\hat{C}$, let us
define $G(t)$ as the number of packets which are inevitably lost up
to time $t$, $I(t)$ as the total number of received innovative
symbols up to $t$. Note that not all innovative symbols can be used
for decoding the original packets. For example, in
Table~\ref{tab:MWNC}, the symbol received at $t=8$ is useless at the
end of time $t=12$ since $p_3$ cannot be decoded ever since. We
define $D(t)$ as the number of discarded symbols up to time $t$.
Then $I(t)-D(t)$ represents the received innovative symbols that
contain the information for the packets covered by the coding window
up to time $t$ (except for the lost $G(t)$ packets).

For a MWNC's receiver with capacity $\hat{C}$, let us define a
particle on $\mathcal{R}^1$ with its position at time $t$ given by
\begin{equation}\label{eq:particle}
S(t)=V\times t-G(t)-(I(t)-D(t)).
\end{equation}

We have the following results regarding the decoding and loss events
for MWNC.

\begin{lemma}\label{lemma1}
Decoding event occurs at time $t$ if and only if $S(t)\le 0$ at the
end of this time slot. All the packets from the last foremost
decoded (or lost) packet to the head of current window will be
decoded.
\end{lemma}

\begin{proof}
Decoding event occurs at the moment when the coding coefficient
matrix in buffer is full rank. In this case, the number of
innovative symbols in buffer must be as many as the number of
packets covered by the window by time $t$ except for the lost ones,
i.e., $\lceil V\times t\rceil - G(t) $. Therefore, we have
$(I(t)-D(t)) = \lceil V\times t\rceil-G(t)$, i.e., $S(t)\leq 0$.
\end{proof}

\begin{lemma}\label{lemma2}
Packet loss event occurs at time $t$ if and only if $S(t)>W-V$ at
the end of $t$. Moreover, all the un-decoded packets before the tail
of the window are lost.
\end{lemma}

\begin{proof}
The last packet in the coding window moves to $\lceil V\times
(t+1)-W+1\rceil$ at time $t+1$. So if the packet right before
$\lceil V\times (t+1)\rceil-W$ has not been ``seen'' (i.e., a symbol
contains this packet has not been received before), then all the
un-decoded packets before the window will get lost forever. In other
words, $\lceil V\times (t+1)\rceil-W>G(t)+((I(t)-D(t))$, which gets
$S(t)>W-V$.
\end{proof}

\begin{lemma}\label{lemma3}
If a new  packet is lost at time $t$, the number of newly discarded
symbols in buffer must be exactly one less than the number of
packets just get lost, i.e.,
\begin{equation}
\Delta D(t)=\Delta G(t)-1.
\end{equation}
\end{lemma}

\begin{proof}
Since the coding window moves constantly, if the packet $\lceil
V\times (t+1)\rceil-W-1$ has not been ``seen'' at time $t$, then the
packet should have been decided as lost in the last time slot, which
contradicts to the assumption.
\end{proof}

\begin{figure}[t]
\centering
\includegraphics[width=\figsize]{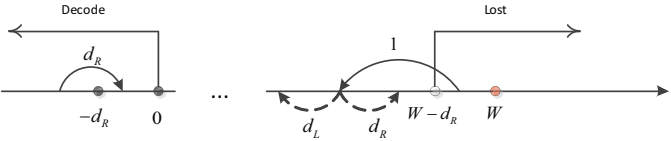}
  \caption{Random walk model for MWNCast.}
  \label{fig:random walk}
\end{figure}

Accordingly, for the specified receiver, MWNC can be modeled as a
one-dimensional random walks \cite{Cox1965} (see Fig.
\ref{fig:random walk}). The random walk has two reflecting barriers
at $-V$ and $W-V$ corresponding to the decoding and packet loss
events respectively, where $S(t)$ in \eqref{eq:particle} corresponds
to the position of the random walk at time $t$. Let $X$ denote the
step size of the random walk, which is a random variable with the
following density function:
\begin{equation}\label{eq:random_walk_step}
f(x)=\hat{C}\delta (x+d_L)+(1-\hat{C})\delta(x-d_R)
\end{equation}
where $d_L=1-V$ and $d_R=V$. Moreover, the mean and variance of a
step are denoted as $\mu=V-C$ and $\sigma^2 = Var(X) =
\hat{C}(1-\hat{C})$.

The following theorem specifies the behavior of the random walk
representing the specified receiver:
\begin{theorem}\label{theorem1}
At time $t$, if the particle crosses the left barrier $-V$, it will
be reflected rightward for a distance of $d_R$ in the next time
slot. If the particle crosses the right barrier $W-V$, it will be
immediately bounced back for a distance of $1$. Otherwise, the
particle will make a random move according to
\eqref{eq:random_walk_step}. \end{theorem}

\begin{proof}
Firstly, notice that if and only if $S(t-1)>-V$, the received symbol
at time $t$ is innovative. That is, the window's foremost packet
$\lceil V\times t\rceil$ is informative to the receiver since
$V(t-1)-G(t-1)-(I(t-1)-D(t-1))>-V$.

Therefore, when the particle does not cross the two barriers, its
position at time $t$ relative to the last slot $S(t)=S(t-1)+V-\Delta
I(t)$ depends on whether a symbol is received successfully, which
follows the the step function defined in
\eqref{eq:random_walk_step}. If the particle just crosses the left
barrier $-V$, the received symbol contains no new information. Thus
$I(t)=I(t-1)$ and the particle moves rightward definitely. If the
particle crosses the right barrier $W-V$, it is indicated in Lemma
\ref{lemma3} that it will be bounced back instantly by a distance of
$1$.
\end{proof}

\subsection{Reliability Analysis}
The reliability analysis is complicated in MWNCast since the
second-hop receivers might suffer from larger packet loss ratio.
However, it is easy to see that the proposed stochastic scheduling
policy converges to the required proportional of time for each relay
set in a long run and consequently the information difference
between BS and the relays should not be large. In addition, the
symbols to be transmitted by the relays are generated randomly, so
that the probability that they contain innovative information to the
second-hop receivers is greatly increased. Therefore, without
differentiating the relays and receivers, we assume a specified
client receives information on a channel of equivalent capacity
$\hat{C}$. Based on this assumption, the packet loss ratio for both
the relays and the receivers can be derived.

\begin{figure}[t]
\centering
\includegraphics[width=2in]{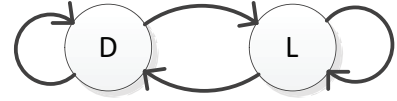}
  \caption{Point process model for decoding and loss events.}
  \label{fig:point process}
\end{figure}

We can model MWNCast as a two-state point process \cite{Cox1965} as
shown in Fig. \ref{fig:point process}, which corresponds to the
``Decode'' (D) and ``Loss'' (L) events, respectively. Specifically,
if the ``Decode'' event occurs at some time, with probability
$P_{DD}$ it will return the same state after a random time interval
$T_{DD}$, and with probability $P_{DL}=1-P_{DD}$ it will make a
transition to the ``Loss'' states after a time interval $T_{DL}$.
Similarly, we can define $ P_{LL}, P_{LD}$ and $T_{LL}, T_{LD}$ as
the transition probabilities and transition time for the ``Loss''
state. These quantities can be derived using the random walk and
point process theories as follows.

Firstly, let us define $G(\theta)$ as the moment generating function
of $X$, which is the two-sided Laplace transform of the step
function $f(x)$ defined in \eqref{eq:random_walk_step}, that is,
\begin{equation}
G(\theta) = E[e^{-\theta X}] = \hat{C} e^{-\theta
(1-V)}+(1-\hat{C})e^{\theta V}.
\end{equation}

From the property of moment generating function, we know that: (i)
$G(\theta)$ is a convex function; (ii) If $E[X] \neq 0$, there are
two roots for the equation $G(\theta) = 1$, one is $\theta = 0$, the
other is $\theta = \theta_0$ who has the same sign as $\mu$.

Let $-B$ and $A$ ($A, B>0$) denote two absorbing barriers for the
random walk starting at the origin, we can define the stopping time
$N$ as
\begin{equation}\label{eq:stopping_time}
N = \min \{n: S(n) \leq -B \:\mbox{or}\: S(n) \geq A\},
\end{equation}
which is the number of steps to cross one of the barriers starting
from the origin.

We can define the moment generating function $G_N(\theta)$ with
respect to $N$ and $S(N)$, that is,
 \begin{equation}\label{eq:G_N}
 G_N(\theta) = E[e^{-\theta S(N)}s^N].
 \end{equation}

Suppose that we set $s = G(\theta)^{-1}$, then we have $G_N(\theta)
= E[e^{-\theta S(N)}G(\theta)^{-N}]$. For this equation, we can find
$\theta = \theta_0$ such that $G(\theta) = 1$, then it is easy to
verify that $e^{-\theta S(n)}G(\theta)^{-n}$ is a martingale with
mean 1 since it is the product of independent unit mean random
variables. According to the martingale stopping theorem (Theorem
6.2.2 in \cite{ross1996}), we can obtain:
\begin{equation}\label{eq:martingale_stopping_theorem}
E[e^{-\theta S(N)}] = 1.
\end{equation}

Since the events of $S(N) \leq -B$ and $S(N)\geq A$ are independent,
from \eqref{eq:stopping_time} and
\eqref{eq:martingale_stopping_theorem}, we have: {\small
\begin{equation}
E[e^{-\theta S(N)}|S(N)\geq A]P_A + E[e^{-\theta S(N)}|S(N)\leq
-B]P_{-B} = 1.
\end{equation}
} For the absorbing states $A$ and $-B$ , we can get the following
approximations: {\small
\[E[e^{-\theta S(N)}|S(N)\geq A] \simeq e^{-\theta A},\quad  E[e^{-\theta S(N)}|S(N)\leq -B] \simeq e^{\theta
B}.
\]
} Substituting these two approximation equations into
\eqref{eq:martingale_stopping_theorem}, and using the fact that $P_A
+ P_{-B} =1$, we can get the probabilities of absorption at $A$ and
$-B$ as
\begin{equation}\label{eq:absorption_prob}
P_A\simeq \frac{1-e^{\theta_0 B}}{e^{-\theta_0 A}-e^{\theta_0 B}},
\quad P_{-B}\simeq \frac{-1+e^{-\theta_0 A}}{e^{-\theta_0
A}-e^{\theta_0 B}},
\end{equation}
where $\theta_0$ is the non-zero root of the equation $G(\theta)=1$.

To derive the distribution for $N$, let $\lambda_1(s)$ and
$\lambda_2(s)$ denote two real roots of the equation $G(\theta) =
1/s$. Then from \eqref{eq:G_N}, we can obtain a different expression
of \eqref{eq:martingale_stopping_theorem} with respect to $N$:
\begin{equation}
E[e^{-\lambda_1(s)S(N)}s^N] = 1,\: E[e^{-\lambda_2(s)S(N)}s^N] = 1.
\end{equation}

Using the approximation $S(N) \simeq A$ when $S(N) \geq A$, and
$S(N) \simeq -B$ when $S(N) \leq -B$, we have:
\begin{equation}\label{eq:EA_EB}
P_Ae^{-\lambda_i(s)A}E_A(s^N)+P_{-B}e^{\lambda_i(s)B}E_{-B}(s^N)=1, i=1,2, \\
\end{equation}
where $E_A$ and $E_{-B}$ denote the conditional expectations at $A$
and $-B$, respectively. Using $P_A$ and $P_{-B}$ given by
\eqref{eq:absorption_prob}, we can obtain $E_A(s^N)$ and
$E_{-B}(s^N)$ from \eqref{eq:EA_EB}. Then we have the moment
generating function for $N$ as:
\begin{equation}\label{eq:S(N)}
E[s^N] = P_A E_A(s^N) + P_{-B}E_{-B}(s^N).
\end{equation}

By differentiating Eq. \eqref{eq:S(N)} with respect to $s$, we can
obtain the first and second moments of $N$ respectively.

To derive the packet loss ratio, we assume the particle is always
located at the largest possible position after an event, which gives
the upper bound for the loss probability. By Theorem \ref{theorem1},
after an ``D'' event, the particle's maximum position is at $d_R$;
After a ``L'' event, the particle must get back to at most $W-1$
before making a random move. Therefore, the transition probabilities
$P_{DD},P_{DL},P_{LD}$ and $P_{LL}$ between these two events, and the
expected transition time $T_{DD},T_{DL},T_{LD}$ and $T_{LL}$ can be
derived using \eqref{eq:absorption_prob} and \eqref{eq:S(N)}
respectively assuming the particle starting from the corresponding
position.

The equilibrium distribution of the embedded Markov chain for
two-state point process can be given by
$(\pi_D,\pi_L)=(\frac{P_{LD}}{P_{DL}+P_{LD}},\frac{P_{DL}}{P_{DL}+P_{LD}})$.
Let us define $T = \pi_D
P_{DL}T_{DL}+\pi_LP_{LL}T_{LL}+\pi_DP_{DD}T_{DD}+\pi_LP_{LD}T_{LD}$,
then the proportion of time passed from states ``L'' and ``D'' to
state ``L'' are given by ${\pi_D P_{DL} T_{DL}}/{T}$ and $\pi_L
P_{LL}T_{LL} /{T}$, respectively.

Note that MWNC has two packet loss scenairo depending on the
previous event, ``L'' to ``L'' and ``D'' to ``L''. If it is from
``L'' to ``L'', all the packets covered by the window will get lost,
so the number of lost packets should be proportional to the
transition time $T_{LL}$. If it is from ``D'' to ``L'', then the
lost packets should be the number of packets covered from the time
of previous event minus $W$ since only the packets behind the window
will get lost. Therefore, the overall packet loss probability can be
obtained as:
\begin{equation}\label{eq:pkt_loss}
P_{loss} = \frac{\pi_DP_{DL}(T_{DL}-W/V)+\pi_LP_{LL}T_{LL}}{T}.
\end{equation}




\subsection{Decoding delay analysis}
For a receiver in MWNCast, the delay for receiving a packet is composed of two parts: the queueing delay and
the decoding delay. The queueing delay can be analyzed using the
similar procedure in \cite{Kumar2008}. In the following, we will focus on
the decoding delay and consider a saturated system in which the
source always has packets to transmitted.

Attributed to the stochastic scheduling policy, any client (a relay
or a receiver) in the network can be approximately considered as
connected to the information source by MWNC on a channel with
equivalent capacity $\hat{C}$. If the window size is sufficiently
large, the packet loss probability is negligible. In this case, we
can derive an upper bound for the average decoding delay assuming
that $W\rightarrow \infty$, whereby the random walk is simplified to
a single left barrier at $0$ with a starting point $d_R$. The
barrier at $0$ indicates the moment of decoding. Assume the particle
always reflects back to the largest possible position $d_R$ after a
decoding event. According to Eq. \eqref{eq:S(N)}, when $A$
approaches to infinity and $B$ is set $d_R$, we have $P_{-B}=1$ and
$P_A=0$. The first two moments of $N$ are derived from
\eqref{eq:S(N)} as follows:

\begin{equation}\label{eq:decoding_delay}
E[N] \simeq -\frac{d_R}{\mu}, E[N^2] \simeq
\frac{{d_R}^2\mu-\sigma^2 d_R}{\mu^3}.
\end{equation}

The {\it decoding delay} for a packet is defined as the time
duration from the moment that it is encoded to the time that it is
decoded. We can model the decoding process as a renewal process.
Then the sum of decoding delay for the packet in a renewal period $N_i$
is given by
\[ D(N_i) \simeq  \frac{N_i}{2}\cdot N_iV,  \]
where $N_i/2$ is the average decoding delay for a packet, $N_i V$ is
the average number symbols transmitted during this time period.

By the theory of renewal reward process and the definition of
average decoding delay, we have
\begin{equation}\label{eq:average delay}
D = \lim\limits_{t\to \infty}\frac{1}{t\cdot
V}\sum\limits_{i=1}^{\infty}D(N_i)=\frac{E(D)}{E(N)\cdot V} =\frac{1
}{2}\frac{E[N^2]}{E[N]}.
\end{equation}

From \eqref{eq:decoding_delay} and \eqref{eq:average delay}, we
can prove that $D$ approaches to $O(\frac{1}{(1-\rho)^2})$
asymptotically, where $\rho={V}/{\hat{C}}$ is the traffic intensity of
the receiver.

\subsection{Decoding Complexity}
In this part, we analyze the decoding complexity of MWNC assuming the window size $W$ is sufficiently large such that the packet loss can be negligible.

The decoding complexity of MWNC is composed of two parts: the forward
elimination and the backward substitution. Suppose that a receiver has just decoded all packets
up to $\lceil V\times t_0\rceil$ at time $t_0$, and $t_1,t_2,...,t_{k-1}$ denote the time instances that the receiver receives a set of $k-1$ encoded symbols but cannot decode them, until at time $t_k$ it receives the $k$th symbol and is able to decode all the received symbols. In the following, we use $s_j$ to denote the symbol received at time $t_j (j=1,\cdots, k)$.

\begin{lemma}\label{lemma4}
The number of nonzero entries in the $j^{th}$ symbol $s_j$ after the
forward elimination is $\lceil S(t_j)\rceil+1$.
\end{lemma}

\begin{proof}
The forward elimination can assure that all the previously ``seen''
packets can be reduced from the newly received symbol. At time
$t_j$, the receiver has received $I(t_j)-D(t_j)$ useful symbols.
Hence, except for $G(t_j)$ inevitably lost ones, all the packets up
to $G(t_j)+I(t_j)-D(t_j)$ have been ``seen'' even they may not be
all decoded. The new symbol generated in the $t_j$'s coding window
can be reduced at the corresponding positions except for the last
one $G(t_j)+I(t_j)-D(t_j)$, which becomes ``seen'' for the sake of
the symbol received at $t_j$. Hence, there are $\lceil V\times
t_j\rceil-G(t_j)-(I(t_j)-D(t_j))+1$ nonzero entries are left, which
equals to $\lceil S(t_j) \rceil$ according to Eq.
\eqref{eq:particle}.
\end{proof}

\begin{lemma}\label{lemma5}
The number of arithmetic operations required for forward elimination
of the $j^{th}$ symbol $s_j$ is $W-\lceil
S(t_j)\rceil-j+\sum_{i=1}^{i=j-1} (\lceil S(t_i)\rceil+1)$ when
$j+\lceil(S(t_j)\rceil<W$, and
$\sum_{i=j-W+1+\lceil(S(t_j)\rceil}^{i=j-1} (\lceil S(t_i)\rceil+1)$
otherwise.
\end{lemma}

\begin{proof}
The packets from $\lceil t_j\times V\rceil-W+1$ to $\lceil t_j\times
V\rceil$ are used to generate the specific symbol. Among them, the
receiver may have decoded a number of packets. Since as Lemma
\ref{lemma4} suggests, the number of nonzero entries after
elimination is $\lceil(S(t_j)\rceil+1$ and there are $j-1$
previously eliminated results, the receiver only has a decoded
intersection with the coding window if $j+\lceil(S(t_j)\rceil<W$. In
this case, the first $W-\lceil S(t_j)\rceil-j$ packets in the window
are already decoded by the receiver, so the same number of
calculations are needed to eliminate these entries. Then, the
previously reduced symbols ($s_i,i<j$) can be used to eliminate the
corresponding entries, which takes $\lceil S(t_i)\rceil+1$
operations for each symbol $s_i$ according to Lemma \ref{lemma4}.
When $j+\lceil(S(t_j)\rceil\ge W$, there are no decoded packets in
the coding window, thus the receiver can only use the existing
eliminated results to reduce the symbol. The elimination for the
$W-\lceil(S(t_j)\rceil-1$ positions takes
$\sum_{i=j-W+1+\lceil(S(t_j)\rceil}^{i=j-1} (\lceil S(t_i)\rceil+1)$
calculations.
\end{proof}

\begin{theorem}\label{theorem2}
Given the target throughput $V$, MWNC can achieve the optimal
decoding complexity of $O(W)$.
\end{theorem}

\begin{proof}
A network coding symbol is encoded with $W$ packets, a receiver needs at least $W-1$ operation to decode the original information, so a trivial lower bound for the decoding complexity is $O(W)$. Therefore, it is sufficient
to prove that, for every receiver in the network, the decoding
complexity of MWNC is upper bounded by $O(W)$.

According to Lemma \ref{lemma5}, an obvious upper bound of
complexity for the $j^{th}$ symbol's elimination procedure is
$W+\sum_{i=1}^{i=j-1} (\lceil S(t_i)\rceil+1)$\footnote{For $j=1$,
the first symbol certainly needs no more than $W$ operations, thus
without abusing the notation, we assume $\sum_{i=1}^{i=0} (\lceil
S(t_i)\rceil+1)=0$.}. Thus, the total number of computations for the
forward elimination of $k$ packets is upper bounded by
$\sum_{j=1}^{j=k} (W+\sum_{i=1}^{i=j-1} (\lceil S(t_i)\rceil+1))$.
The total number of computations for backward substitution is
$\sum_{j=1}^{j=k} (\lceil S(t_j)\rceil+1)$, so the overall
complexity $\Omega(k)$ for decoding $k$ packets is bounded by:
\begin{equation}\label{eq:k complexity}
\begin{split}
\Omega(k) & \le Wk+\frac{(k+1)k}{2}+\sum_{j=1}^{k}\sum_{i=1}^{j}\lceil
S(t_i)\rceil  \\
&\le \frac{(k+3 )k}{2}W+\frac{(k+1)k}{2},
\end{split}
\end{equation}
where the second inequality is valid because  $\lceil
S(t_i)\rceil\le W$.

Note that as discussed in previous subsection, the decoding event
occurs after a random time duration of $N$, so $k=\lceil
V(t+N)\rceil-\lceil Vt\rceil\simeq VN$. According to the renewal
reward process and \eqref{eq:k complexity}, the average decoding
complexity can be obtained as follows:
\begin{equation}\label{eq:average complexity}
\Omega \le\frac{E(\Omega(N))}{E(N)\cdot V} =\frac{E(N^2)V +
3E(N)}{2E(N)}W+\frac{E(N^2)V + E(N)}{2E(N)}.
\end{equation}

From \eqref{eq:decoding_delay}, we know that given the throughput,
$E(N)$ and $E(N^2)$ are independent of $W$, therefore, the decoding
complexity is dominated by the window size, which is on the order of
$O(W)$ from \eqref{eq:average complexity}
\end{proof}

\section{Simulation Results}\label{sec:simulation}
In this section, we provide extensive simulation results to compare
the performance of MWNCast with other network coding-based multicast
schemes, and also to illustrate the advantages of MWNC technique. The network topologies in
simulations are generated randomly, whereby the location of all
clients is uniformly distributed around the source. The channel
between any two nodes is assumed to follow the Rayleigh fading channel model. The
coefficients of network coding are generated on a $G(2^8)$ Galois
Field. The time duration for each simulation is $10^5$ time
slots.

\subsection{Throughput and Decoding Delay}

\begin{figure}[t]
\centering
\includegraphics[width=\figsize]{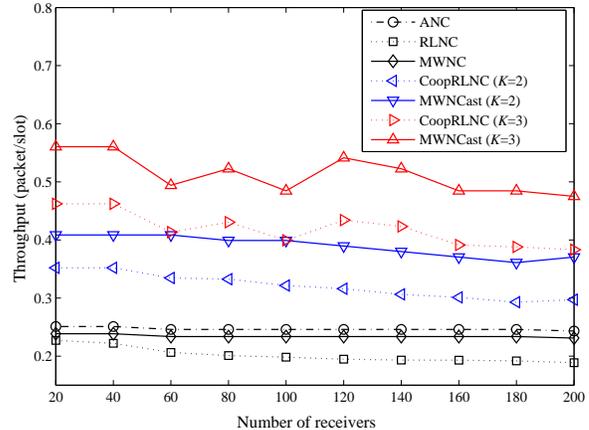}
  \caption{Throughput vs. Network size.}
  \label{fig:network scale}
\end{figure}

In  Fig. \ref{fig:network scale}, we compare the achievable
throughput of MWNCast\footnote{The packet loss probability is
controlled under $10^{-3}$ as will be explained in the next
subsection.} with those of RLNC and ANC under different network
sizes, where the window size of MWNCast is set to $20$.
When the channel number $K=1$, MWNCast reduces to MWNC (simple multicast without cooperation). We can see that the achieved throughput of MWNC is
close to ANC (which is known to be throughput-optimal), but it
outperforms RLNC under all network conditions. It also can be seen
that the throughput of RLNC decreases as the network size increases,
which has been discussed in the literature
\cite{Swapna2010}. When multiple channels are available, we provide the simulation for RLNC with the same relay scheduling
strategy as in MWNCast for fair comparison. It is observed that for
$K=2$ and $K=3$, MWNCast preserves its superiority over cooperative RLNC (denoted by CoopRLNC in Fig. \ref{fig:network scale}) with the average performance gain of $21.5\%$ and $22.5\%$, respectively. Therefore, we can
see that by taking advantage of moving window network coding and
cooperation, MWNCast is effective in improving  the system
throughput.

\begin{figure}[t]
\centering
\includegraphics[width=\figsize]{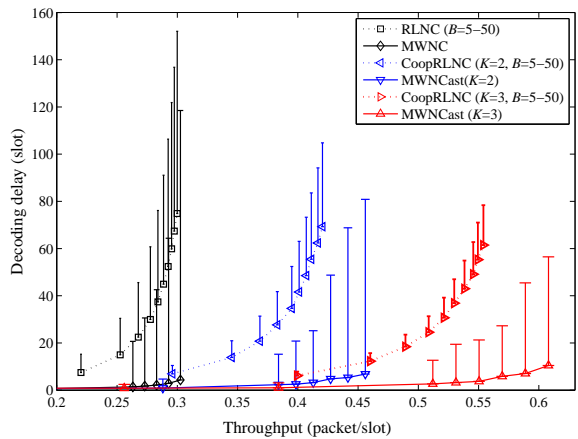}
  \caption{Decoding delay vs. Throughput.}
  \label{fig:delay throughput}
\end{figure}

In Fig. \ref{fig:delay throughput}, we study the tradeoff between the
decoding delay and throughput for a network with 100 nodes. Note that the
result of ANC is not included since the decoding delay of the
receivers with poorer channel condition increases with the
simulation time, which is unfair for comparison. In this figure, the
lines and bars show the average and maximum decoding delays of all
receivers respectively for each scheme. It is noteworthy that, to
maintain a given throughput, with the newly proposed MWNC-based
scheme, the average decoding delay for a successful decoded packet
is much lower than that with the RLNC-based scheme. The reason is
that the decoding opportunity with MWNC exists in each time slot,
but with RLNC, it cannot decode until receiving the full block of
packets.

\subsection{Reliability}
\begin{figure}[t]
\centering
\includegraphics[width=\figsize]{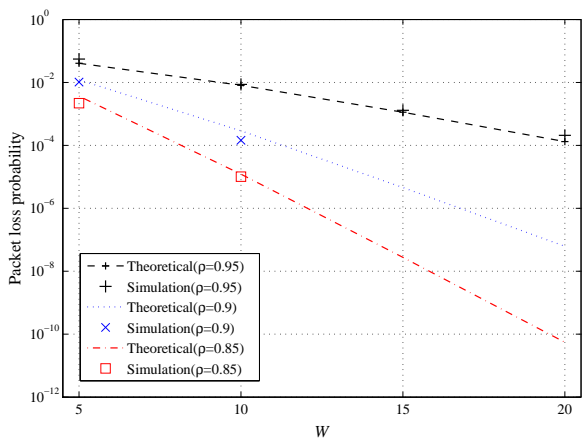}
  \caption{Packet loss probability vs. Window size $W$.}
  \label{fig:loss ratio}
\end{figure}

In Fig. \ref{fig:loss ratio}, we study the effect of window size $W$
on the packet loss ratios of MWNCast under different system traffic load
conditions \footnote{System traffic load is defined as $\rho=V/C^*$.} when the
channel number $K=2$. The packet loss counted in the simulation can
occur at any clients in the network. From the figure, it can be seen
that the theoretical results match well with the simulation results.
Also, we notice that, the packet loss probability drops almost
exponentially with the increase of window size. Moreover, we can see
the requirement of packet loss probability of $10^{-3}$ can be
satisfied with  $W=20$ even when the traffic load is as high as
$0.9$.

If a certain degree of packet loss can be tolerated (i.e.
$10^{-3}\sim10^{-1}$), we study the relationship between the average
decoding delay and packet loss probability by adjusting the window
size. We consider a specific receiver with the packet erasure
probability equals to $0.3$. As shown in Fig. \ref{fig:reliaiblity
delay}, for a given traffic load, the packet loss probability
increases with the decrease of window size, but the average decoding
delay for the packets not lost also gets smaller. As discussed in
Lemma \ref{lemma2}, this is because packet loss is inevitable when
and only when the the last unseen packet is just the one before the
lattermost packet in the window. Such packets, if not lost, will
encumber the decoding of the following packets, leading to larger
overall decoding delay. Therefore, with a smaller coding window
size, MWNCast acts as a delay filter to force the receiver to drop
the packets which may lead to the degradation of the overall delay
performance.

\subsection{Decoding Complexity}

\begin{figure}[t]
\centering
\includegraphics[width=\figsize]{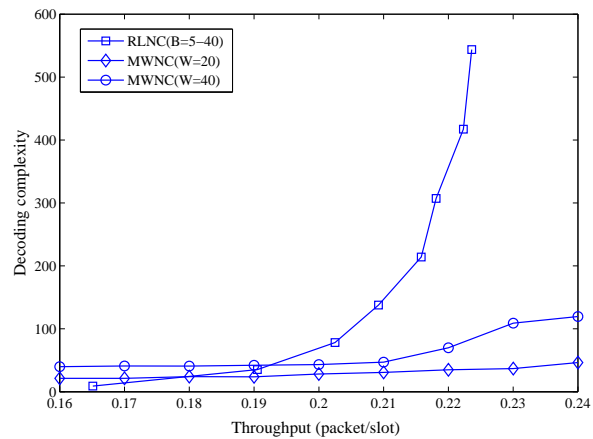}
  \caption{Decoding complexity vs. Throughput.}
  \label{fig:complexity}
\end{figure}

In Fig. \ref{fig:complexity}, we compare the decoding complexity of
MWNC and RLNC. It can be seen that the decoding complexity of MWNC
is much smaller than that of RLNC. It is found that to decode RLNC,
the forward elimination is the dominating part (i.e., $O(W^3)$) as
$W$ increases and hence the average complexity for decoding an
original packet is $O(W^2)$. Nevertheless, in order to achieve
higher throughput, larger batch sizes have to be used, so the
decoding complexity of RLNC increases dramatically with the growing
of throughput. For MWNC, the decoding coefficient matrix of the
receiver is sparse since it has smaller number of non-zero items. As
a result, the decoding complexity consumed by Gauss Elimination is
significantly reduced (proportional to the window size) and has
little dependency on the throughput.


\begin{figure}[t]
\centering
\includegraphics[width=\figsize]{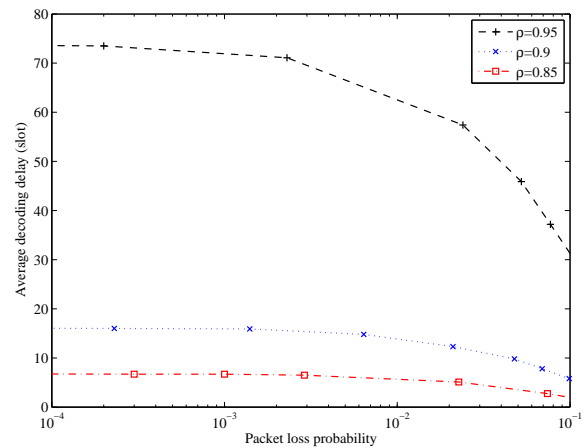}
  \caption{Decoding delay vs. Reliability.}
  \label{fig:reliaiblity delay}
\end{figure}

\section{Conclusions}\label{sec:conclusion}
In this paper, we proposed MWNC, a novel network coding scheme that
has smaller decoding delay, lower complexity and no need for
feedback from the receivers. Based on this technique, we further
present the MWNCast protocol to address the bottleneck problem in
wireless multicast through cooperative relays. Theoretical analysis
shows that the proposed schemes can approach the network capacity
with the packet loss probability dropping almost exponentially with
the increase of window size, the average decoding delay  is on the
order of $O(\frac{1}{(1-\rho)^2})$ and the decoding complexity is on
the order $O(W)$. Simulation results are provided to validate the
theoretical results and compare the performance of the proposed
schemes with the existing solutions.

{\footnotesize
\bibliography{reference}

\begin{thebibliography}{10}

\bibitem{Wu2012Globecom}
F.~Wu, C.~Hua, H.~Shan, and A.~Huang, ``Mwncast: cooperative multicast based on
  moving window network coding,'' in {\em IEEE GLOBECOM'12}, pp.~1--7, 2012.

\bibitem{2006two}
G.~Papadopoulos, G.~Koltsidas, and F.~Pavlidou, ``Two hybrid arq algorithms for
  reliable multicast communications in umts networks,'' {\em IEEE
  Communications Letters}, vol.~10, no.~4, pp.~260--262, 2006.

\bibitem{zhang2006}
X.~Zhang and Q.~Du, ``Adaptive low-complexity erasure-correcting code-based
  protocols for qos-driven mobile multicast services over wireless networks,''
  {\em IEEE Transactions on Vehicular Technology}, vol.~55, no.~5,
  pp.~1633--1647, 2006.

\bibitem{luby2002}
M.~Luby, ``{LT} codes,'' in {\em Proceedings of the IEEE Symposium on the
  Foundations of Computer Science}, pp.~271--280, 2002.

\bibitem{Shokrollahi2006}
A.~Shokrollahi, ``Raptor codes,'' {\em IEEE/ACM Trans. on Networking}, vol.~14,
  pp.~2551--2567, June 2006.

\bibitem{ho2004}
T.~Ho, {\em Networking from a network coding perspective}.
\newblock PhD thesis, MIT, 2004.

\bibitem{alay2009}
O.~Alay, P.~Liu, Z.~Guo, L.~Wang, Y.~Wang, E.~Erkip, and S.~Panwar,
  ``Cooperative layered video multicast using randomized distributed space time
  codes,'' in {\em IEEE INFOCOM Workshops}, pp.~1--6, 2009.

\bibitem{rong2010}
B.~Rong and A.~Hafid, ``Cooperative multicast for mobile iptv over wireless
  mesh networks: The relay-selection study,'' {\em IEEE Transactions on
  Vehicular Technology}, vol.~59, no.~5, pp.~2207--2218, 2010.

\bibitem{zhao2010}
H.~Zhao and W.~Su, ``Cooperative wireless multicast: performance analysis and
  power/location optimization,'' {\em IEEE Transactions on Wireless
  Communications}, vol.~9, no.~6, pp.~2088--2100, 2010.

\bibitem{fan2009}
P.~Fan, C.~Zhi, C.~Wei, and K.~Ben~Letaief, ``Reliable relay assisted wireless
  multicast using network coding,'' {\em IEEE Journal on Selected Areas in
  Communications}, vol.~27, no.~5, pp.~749--762, 2009.

\bibitem{jin2009}
J.~Jin and B.~Li, ``Cooperative multicast scheduling with random network coding
  in wimax,'' in {\em 17th International Workshop on Quality of Service
  (IWQoS'09)}, pp.~1--9, 2009.

\bibitem{fanous2010}
A.~Fanous and A.~Ephremides, ``Network-level cooperative protocols for wireless
  multicasting: Stable throughput analysis and use of network coding,'' in {\em
  IEEE Information Theory Workshop (ITW)}, pp.~1--5, 2010.

\bibitem{Pac}
D.~Koutsonikolas, Y.~C. Hu, and C.-C. Wang, ``Pacifier: High-throughput,
  reliable multicast without crying babies in wireless mesh networks,'' {\em
  IEEE/ACM Transactions on Networking}, vol.~PP, no.~99, p.~1, 2011.

\bibitem{Swapna2010}
B.~Swapna, A.~Eryilmaz, and N.~Shroff, ``Throughput-delay analysis of random
  linear network coding for wireless broadcasting,'' in {\em NetCod'10}, June
  2010.

\bibitem{More}
S.~Chachulski, M.~Jennings, S.~Katti, and D.~Katabi, ``Trading structure for
  randomness in wireless opportunistic routing,'' {\em SIGCOMM Comput. Commun.
  Rev.}, vol.~37, pp.~169--180, Aug. 2007.

\bibitem{Padhye2005}
J.~Padhye, S.~Agarwal, V.~N. Padmanabhan, L.~Qiu, A.~Rao, and B.~Zill,
  ``Estimation of link interference in static multi-hop wireless networks,'' in
  {\em Proceedings of the 5th ACM SIGCOMM conference on Internet Measurement
  (IMC'05)}, pp.~28--28, 2005.

\bibitem{Reis2006}
C.~Reis, R.~Mahajan, M.~Rodrig, D.~Wetherall, and J.~Zahorjan,
  ``Measurement-based models of delivery and interference in static wireless
  networks,'' in {\em SIGCOMM '06}, pp.~51--62, 2006.

\bibitem{Kumar2008}
J.~Kumar~Sundararajan, D.~Shah, and M.~Medard, ``Arq for network coding,'' in
  {\em IEEE ISIT'08}, pp.~1651 --1655, July 2008.

\bibitem{2009minimizing}
W.~Yeow, A.~Hoang, and C.~Tham, ``Minimizing delay for multicast-streaming in
  wireless networks with network coding,'' in {\em IEEE INFOCOM'09},
  pp.~190--198, 2009.

\bibitem{2008online}
J.~Sundararajan, D.~Shah, and M.~M{\'e}dard, ``Online network coding for
  optimal throughput and delay-the three-receiver case,'' in {\em International
  Symposium on Information Theory and Its Applications (ISITA'08)}, pp.~1--6,
  2008.

\bibitem{Barros2009}
J.~Barros, R.~Costa, D.~Munaretto, and J.~Widmer, ``Effective delay control in
  online network coding,'' in {\em IEEE INFOCOM'09}, pp.~208 --216, April 2009.

\bibitem{sorour2010}
S.~Sorour and S.~Valaee, ``Minimum broadcast decoding delay for generalized
  instantly decodable network coding,'' in {\em IEEE GLOBECOM'10}, pp.~1--5,
  2010.

\bibitem{keller2008}
L.~Keller, E.~Drinea, and C.~Pragouli, ``Online broadcasting with network
  coding,'' in {\em Fourth Workshop on Network Coding, Theory and Applications
  (NetCod'08).}, pp.~1--6, 2008.

\bibitem{hou2011}
I.~Hou, P.~Kumar, {\em et~al.}, ``Broadcasting delay-constrained traffic over
  unreliable wireless links with network coding,'' in {\em Proceedings of the
  Twelfth ACM International Symposium on Mobile Ad Hoc Networking and
  Computing}, p.~4, 2011.

\bibitem{Wu2012Pimrc}
F.~Wu, C.~Hua, H.~Shan, and A.~Huang, ``Reliable network coding for minimizing
  decoding delay and feedback overhead in wireless broadcasting,'' in {\em IEEE
  PIMRC'12}, pp.~1--6, 2012.

\bibitem{sundararajan2009network}
J.~Sundararajan, D.~Shah, M.~M{\'e}dard, M.~Mitzenmacher, and J.~Barros,
  ``Network coding meets tcp,'' in {\em IEEE INFOCOM'09}, pp.~280--288, 2009.

\bibitem{lin2010slideor}
Y.~Lin, B.~Liang, and B.~Li, ``Slideor: Online opportunistic network coding in
  wireless mesh networks,'' in {\em IEEE INFOCOM'10}, pp.~1--5, 2010.

\bibitem{feedback}
G.~Wang, X.~Zhao, and X.~Dai, ``On efficient network coding scheme with lossy
  and delayed feedback,'' in {\em 2011 International Symposium on Network
  Coding (NetCod)}, pp.~1 --6, july 2011.

\bibitem{Cohen2008}
R.~Cohen and L.~Katzir, ``The generalized maximum coverage problem,'' {\em Inf.
  Process. Letter}, vol.~108, pp.~15--22, September 2008.

\bibitem{Cox1965}
D.R.Cox and H.~Miller, {\em Theory of Stochastic Processes}.
\newblock Methuen, London, 1965.

\bibitem{ross1996}
S.~M. Ross, {\em Stochastic Processes}.
\newblock John Wiley \& Sons, 1996.

\end{thebibliography}
\bibliographystyle{ieeetr}
}

\end{document}